\documentclass[10pt]{iopart}
\usepackage{iopams}
\usepackage{graphicx}
\usepackage{cite}
\begin{document}
\title{Tunable Thermoelectric Superconducting Heat Pipe and Diode}
\author{F. Antola$^1$, A. Braggio$^1$, G. De Simoni$^1$, F. Giazotto$^1$}
\address{$^1$ NEST Istituto Nanoscienze-CNR and Scuola Normale Superiore, I-56127 Pisa, Italy}
\ead{filippo.antola@sns.it}
\begin{abstract}
Efficient heat management at cryogenic temperatures is crucial for superconducting quantum technologies. This study demonstrates the controlled manipulation of the heat flow and heat rectification through an asymmetric superconducting tunnel junction. The system exhibits a non-reciprocal behavior, developing a thermoelectric regime exclusively when the electrode with the larger gap is heated. This feature significantly boosts thermal rectification effectively classifying the device as a heat diode. At the same time when operating as a thermoelectric engine, the same device also functions as a heat pipe, expelling heat from the cryogenic environment, minimizing losses at the cold terminal. This dual functionality is inherently passive, and the performance of the heat pipe and the heat diode can be finely adjusted by modifying the external electrical load.\\
      
\end{abstract}

\maketitle
\ioptwocol
\section{Introduction}
Nanoscale thermal transport management \cite{fornieri2017towards,pekola2021colloquium,majidi2024heat} has emerged as a focal point of interest because of its applications in quantum technology, including cooling \cite{nahum1994electronic,giazotto2006opportunities,muhonen2012micrometre,marchegiani2018chip}, energy harvesting \cite{sothmann2014thermoelectric,thierschmann2015three,martin2013nanoengineering,jaliel2019experimental,balduque2024coherent}, and radiation detection \cite{enss2005cryogenic,paolucci2023highly,karimi2024bolometric,subero2023bolometric}. From this perspective, the development of a component that behaves as a heat diode is particularly intriguing. This system enables the control of the magnitude of the heat current based on the direction of the thermal gradient \cite{martinez2015rectification,sanchez2017single,zhang2021thermal,kobayashi2020thermal,chang2006solid,malik2022review,wang2017experimental,bhandari2021thermal,giazotto2020very,khomchenko2022voltage,chatterjee2024quasiparticles,tesser2022heat,iorio2021photonic}. Like all diodes, such a system requires intrinsic non-reciprocity. In this regard, the rectification of electronic heat currents has been discussed in SIS' Josephson junctions due to the nonlinear temperature behavior of the superconducting density of staes \cite{giazotto2005josephson,martinez2013efficient}.
Recently, it has been demonstrated that this type of system can develop a thermoelectric regime when a non-linear thermal bias is applied to the larger gap electrode \cite{marchegiani2020nonlinear,marchegiani2020superconducting,germanese2022bipolar,hijano2023bipolar,battisti2024bipolar}. This study discusses the implications of the thermoelectric effect on heat transport, with a particular focus on enhancing the reliability and performance of SIS' thermal diodes. We propose the development of a device designed not only to improve rectification performance but also to facilitate the redirection of heat flow away from cryogenic environments. These two applications are closely connected, as the generated thermoelectric power, which enhances the heat current, can be redirected to a remotely located resistor, where it can be dissipated as Joule heating. Alternatively, the electrical power can be redirected to an additional cooling stage further enhancing the overall cooling efficiency. This enables enhanced electronic control over heat transport, which is crucial for the efficiency of cryogenic temperature systems.\\
\section{Model}
\begin{figure}[t!]
    \centering    \includegraphics[width=0.99\columnwidth]{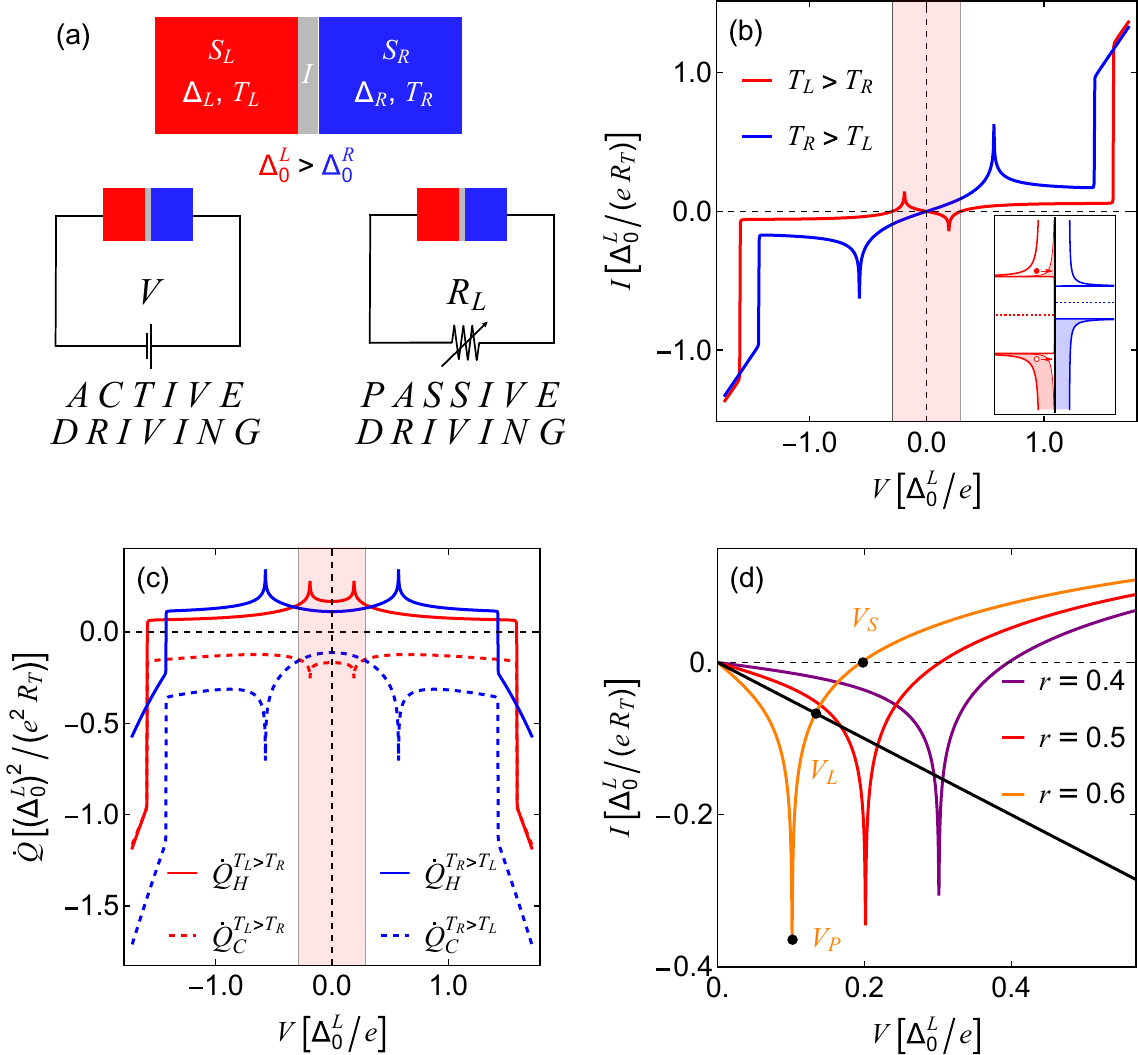}
    \caption{(a) Scheme of the heat-rectifier coupled to a voltage source (left) and to an external resistor (right) (b) $I-V$ with \(r = 0.7\) for \(T_L = 0.6 T_c^L\) and \(T_R = 0.01 T_c^L\) (red), and for the opposite thermal bias \(T_L = 0.01 T_c^L\) and \(T_R = 0.6 T_c^L\) (blue). $T_c^L$ is the critical temperature of the larger gap electrode. The thermoelectric region is shown in pink, while the inset depicts this effect in the energy band diagram. (c) $\dot{Q}-V$ characteristics with the same parameters as in the previous image. $\dot{Q}_H$ refers to the heat current flowing from the hot electrode, while $\dot{Q}_C$ denotes the heat coming to the cold island. The lines' color refers to opposite thermal gradients. (d) Intersection with the load line (black) for \(R_L = 2 R_T\) in the thermoelectric regime for different gap ratios $r = \Delta_R^0/\Delta_L^0$: \(r=0.4\) (orange), \(r=0.5\) (red), and \(r=0.6\) (brown). The temperature is fixed at $T_L=0.8T_c^L$ and $T_R=0.01T_c^L$. The plot also shows the Seebeck voltage $V_S$, corresponding to the open-circuit condition, and the  matching peak \(V_P=|\Delta_L(T_L)-\Delta_R(T_R)|/e\).}
    \label{Figura 1}
\end{figure}

We investigate the thermal and electrical transport behaviors of a superconducting junction as shown in Fig. \ref{Figura 1}(a). The two superconducting islands \(S_\alpha\) (\(\alpha=R, L\)), are characterized by their respective electronic temperatures, \(T_\alpha\), and their temperature-dependent self-consistent energy gaps are very well approximated with \(\Delta_\alpha(T_\alpha)\approx\Re\left[1.746k_BT_c^\alpha\tanh\left({1.74\sqrt{T_c^\alpha/T_\alpha- 1}}\right)\right]\), where $T_c^\alpha$ refer to the critical temperature of the electrode. Our analysis exclusively targets the quasiparticle transport, which can be isolated from the Josephson contribution through a specific junction design or by applying a magnetic field \cite{marchegiani2020superconducting,guarcello2023bipolar}. However, a moderate Josephson effect does not entirely suppress this phenomenon, as reported in Ref.\cite{marchegiani2020phase,germanese2023phase}, promising a phase coherent control to the heat performances of the device.
In our model, we assume the electronic distribution of each electrode to be in equilibrium, following the Fermi-Dirac distribution for quasiparticles \(f_\alpha(E_\alpha, T_\alpha) = (1 + e^{E_\alpha / k_B T_\alpha})^{-1}\), with \(E_\alpha = E - \mu_\alpha\). To calculate the I-V characteristics of the device, we apply a potential difference \(\mu_L - \mu_R = -eV\) across the junction, \textit{i. e.} we adopt a so-called active driving circuital configuration, illustrated in the left side of Fig. \ref{Figura 1}(a). In this framework, the quasiparticle transport of heat and charge flowing out of electrode \(\alpha\), with \(\bar{\alpha}\) labeling the opposite electrode, is described by the following equation \cite{tinkham2004introduction,de2018superconductivity,wehling2014dirac}:
\begin{eqnarray}
 \left(\begin{array}{c}
        \dot{Q}_\alpha\\
        I_\alpha
    \end{array}\right) &=& \frac{1}{e^2 R_T} \int_{-\infty}^{\infty} \left(\begin{array}{c}
        E_\alpha \\ -e
    \end{array}\right) N_\alpha(E_\alpha, T_\alpha) N_{\bar{\alpha}}(E_{\bar{\alpha}}, T_{\bar{\alpha}}) \nonumber \\
    && \times [f_\alpha(E_\alpha, T_\alpha) - f_{\bar{\alpha}}(E_{\bar{\alpha}}, T_{\bar{\alpha}})] \, dE.
\end{eqnarray}
Here, \(R_T\) denotes the normal state resistance of the junction, and 
\begin{equation}
    N_\alpha(E_\alpha, T_\alpha) = \Re\left[\frac{E_\alpha + i\Gamma}{\sqrt{(E_\alpha + i\Gamma)^2 - \Delta_\alpha(T_\alpha)^2}}\right]
\end{equation}
the normalized smeared BCS quasiparticles' Density of States (DOS). The Dynes parameter \(\Gamma\) effectively accounts for inelastic scattering processes in both electrodes. The electron-hole symmetry in the superconducting DOS, namely \(N_\alpha(E_\alpha, T_\alpha) = N_\alpha(-E_\alpha, T_\alpha)\), results in symmetric relationships for heat and reciprocal charge currents as functions of the applied voltage, leading to
\begin{equation}
    \dot{Q}(V, T_L, T_R) =\dot{Q}(-V, T_L, T_R),
\end{equation}
\begin{equation}\label{I rec}
    I(V, T_L, T_R) = -I(-V, T_L, T_R),
\end{equation}
where $\dot{Q}(V, T_L, T_R) = \dot{Q}_R(V, T_L, T_R)$ and $I(V, T_L, T_R) = I_R(V, T_L, T_R)$. When the system is thermoelectric, it exhibits a negative conductance, \(G(V, T_L, T_R) = I(V, T_L, T_R)/V\), which results in a positive thermoelectric power \(\dot{W} = -IV > 0\). Due to the electron-hole symmetry, within the linear response regime, it is impossible to achieve this condition, as indicated by the current reciprocity in Eq. \ref{I rec} \cite{benenti2017fundamental}. Nonetheless, this becomes feasible in systems operating in a non-linear temperature regime, \( T_L - T_R \gtrsim \Delta T_{min}(r)\), and characterized by an energy gap ratio \(r = \Delta_R^0/\Delta_L^0 < 1\). The threshold \(\Delta T_{min}(r)\) can be computed numerically and is also connected to the Dynes parameter $\Gamma$ and always maintains a finite value \cite{marchegiani2020superconducting}.
This situation is illustrated in Fig. \ref{Figura 1}(b), in which we emphasize that the desired bipolar thermoelectric behavior ($IV<0$) requires heating the larger gap superconducting lead. Specifically, in the subgap region, both the $I-V$ characteristics of the system exhibit peaks at the matching point, which arises from the alignment of BCS singularities at \(V_P=\pm|\Delta_L(T_L)-\Delta_R(T_R)|/e\). While in the normal dissipative regime \(I(V_P,T_C,T_H)V_P>0\), in the thermoelectric one \(I(V_P,T_H,T_C)V_P<0\), which is represented by the pink area for the red curve.\\
In the active driving configuration, examination of the heat transport in both scenarios reveals interesting dynamics in the subgap regime $eV\lesssim\Delta_L+\Delta_R$, as we show in Fig. \ref{Figura 1}(c). When the small gap electrode serves as the hot electrode (blue curves), the heat extracted from it (solid line) is lower than that received by the colder electrode (dashed line). This discrepancy is due to the power injected in the junction by the external voltage source, following from the first law of thermodynamics, which requires \(\dot{Q}_R+\dot{Q}_L+IV=0\). Conversely, in the opposite thermal bias, the system may generate bipolar thermoelectricity. In this condition, the power generated in the external circuit ($IV<0$) is obtained at the expense of the heat flowing out from the hotter side (solid). In contrast, the heat in the colder one is reduced (dashed). This is the behavior of a typical thermoelectric device that operates as a thermodynamical engine, producing power dissipated in the external circuit. 
This peculiar thermoelectricity only operates in a range of voltages $|V|\leq V_S$ (pink area in Figs. \ref{Figura 1} (b) and \ref{Figura 1} (c)), where $V_S$ is the Seebeck voltage marking the open-circuit condition \(I(V_S, T_L, T_R)=0\). Increasing the applied voltage further leads the system to recover the dissipative behavior, and the Joule heating warms both leads, resulting in all heat fluxes turning negative.\\
Moreover, the inherent asymmetry between the superconducting gaps naturally facilitates the heat flow from the electrode with the larger gap to the one with the smaller one. However, we anticipate that the generation of thermoelectric power, which may occur when the side with the higher gap is heated, further amplifies the non-reciprocal nature of the heat flow. To harness the full potential of our system, we focus on the circuit configuration with passive driving displayed on the right side of Fig. \ref{Figura 1}(a), where the junction is only closed on a load resistor $R_L$. In this arrangement, the current and the bias are determined by solving:
\begin{equation}\label{Eq Carico}
    I(V_L,T_L,T_R) + \frac{V_L}{R_L} = 0.
\end{equation}
This equation yields a non-zero \(V_L\ne0\) solution exclusively in thermoelectric conditions (\(IV_L < 0\)). Specifically, when the load line intersects the I-V curve within this region, the system can exhibit an odd number of distinct solutions (usually three metastable, referring to solutions with positive differential conductance\cite{germanese2022bipolar}). In particular, the particle-hole symmetric point \(V=0\) is metastable \cite{marchegiani2020nonlinear,germanese2022bipolar,marchegiani2020phase,germanese2023phase}. However, by applying a bias current \(I_B\), it becomes possible to select another thermoelectric metastable solution characterized by \(V=\pm V_L\), with the sign determined by the polarity of \(I_B\). Importantly, once this state is activated by \(I_B\), the system maintains this solution even after \(I_B\) is reset to zero due to the spontaneous particle-hole symmetry mechanism behind the bipolar thermoelectricity\cite{germanese2022bipolar,marchegiani2020nonlinear}. For a specific resistor, the value of $|V_L|$ lies in the range $|V_P|\leq|V_L|\leq|V_S|$, as illustrated in Fig. \ref{Figura 1}(d) for different gap ratios $r$, and by the applied thermal bias. Optimal thermoelectric performance occurs near the matching peak $V_P$ but decreases as it approaches the Seebeck voltage \(V_S\). 
\section{Heat Diode}
In the passive driving configuration, the operating points in the thermoelectric phase determine our system's heat current rectification performance. This allows us to characterize it as a passive thermal diode, which we refer to the stationary condition. \\
To quantify this non-reciprocity we introduce the heat-transport rectification coefficient \(\cal{R}\) as follows:
\begin{equation}
    \mathcal{R}(V_L) = \frac{\dot{Q}^+(V_L) - \dot{Q}^-(0)}{\dot{Q}^-(0)} \times  100,
\end{equation}
where we define
\(\dot{Q}^+(V_L)=\dot{Q}_R(T_L=T_{H}, T_R=T_{C},V_L)\) as the magnitude of the heat current when the electrode with the larger gap is heated and \(\dot{Q}^-(0)=(T_L=T_{C}, T_R=T_{H},V_L=0)\) when the direction of the thermal bias is reversed. We stress that only when the system is thermoelectric a non-null stable solution \(V_L\neq 0\) of Eq.~\ref{Eq Carico} is possible.\\
\begin{figure}[t!]
    \centering       
    \includegraphics[width=\linewidth]{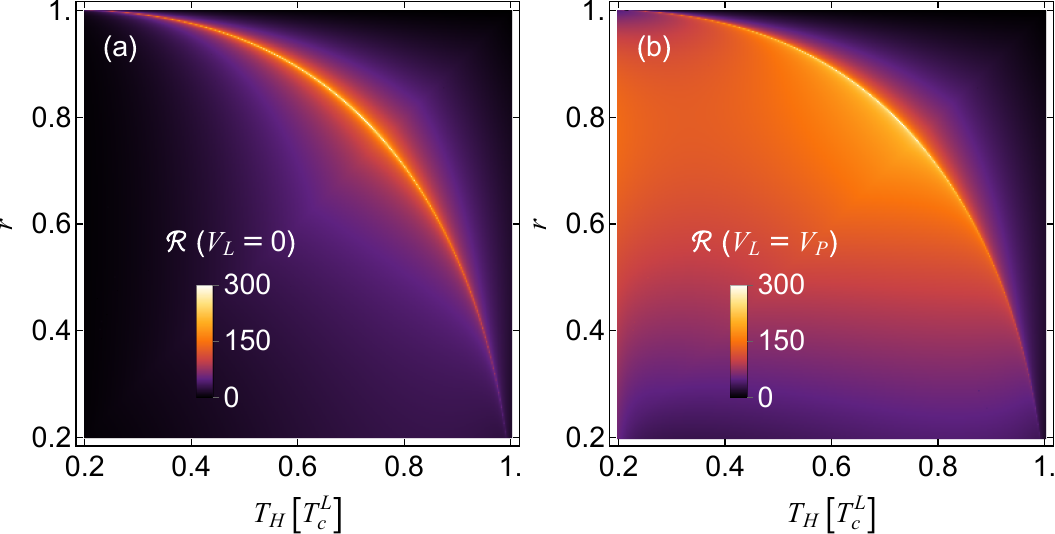}
    
    \caption{Rectification efficiency $\mathcal{R}$ vs $(T_{H},r)$ at $V_L=0$ in (a) and at $V_L=V_P$ in (b). In both cases we set $T_{C}=0.01 T_c^L$.}
    \label{Figura 2}
\end{figure}
Firstly, we start to compare the rectification
at \(V_L=0\), assuming that the SIS' junction remains trapped in the non-thermoelectric metastable phase. This case, which has also been considered before\cite{martinez2013efficient}, could happen when the Josephson coupling is sufficiently strong, shortening the junction that remains trapped at $V_L=0$ not developing any thermoelectricity\cite{marchegiani2020phase,germanese2023phase}. In Fig. \ref{Figura 2}(a), we show the evolution of the rectification by gap asymmetry \(r\) and hot-lead temperature \(T_{H}\), assuming the cold temperature is fixed at $T_C=0.01T_c^L$. We report rectification values up to approximately \(\mathcal{R}_{max}\approx 260\%\), albeit within a constrained parameter space of $T_H$ and $r$. In contrast, when the phenomenon of thermoelectricity is present, it markedly enhances this effect and broadens the applicable conditions, as highlighted in Figure \ref{Figura 2}(b) for \(V_L =V_P\). Indeed, at this operating point, the thermoelectricity and heat extraction from the hot electrode are maximal.
To adopt this operational point, one needs to optimize the value of load resistance. Therefore,  it is worth exploring how the effect remains evident when considering a fixed load resistor, a case that resembles the expected operational use of the device. This persistence is demonstrated in Fig. \ref{Figura 3}(a) (Fig. \ref{Figura 3}(b)) where we consider a fixed resistor $R_L=0.5 R_T$ ($R_L=R_T)$, indicating that substantial heat rectification performance is achievable even without a precise tuning of the load to the parameter variation. In Fig. \ref{Figura 3}(c), we report, as a function of $T_H$ for fixed $r=0.8$, the comparison between the two previous cases and the case of a shortened circuit $R_L=0$, i.e. $V_L=0$. We see that the large values of heat rectification are widened with respect to the resonant value, enhancing its resilience to thermal fluctuations.
Specifically, as we can see in Fig. \ref{Figura 3}(d), choosing a higher \(R_L\) results in a lower slope of the load line. This increases the range over which non-null thermoelectric solutions of Eq. \ref{Eq Carico} can exist.\\
\begin{figure}[t!]
    \centering
   \includegraphics[width=\linewidth]{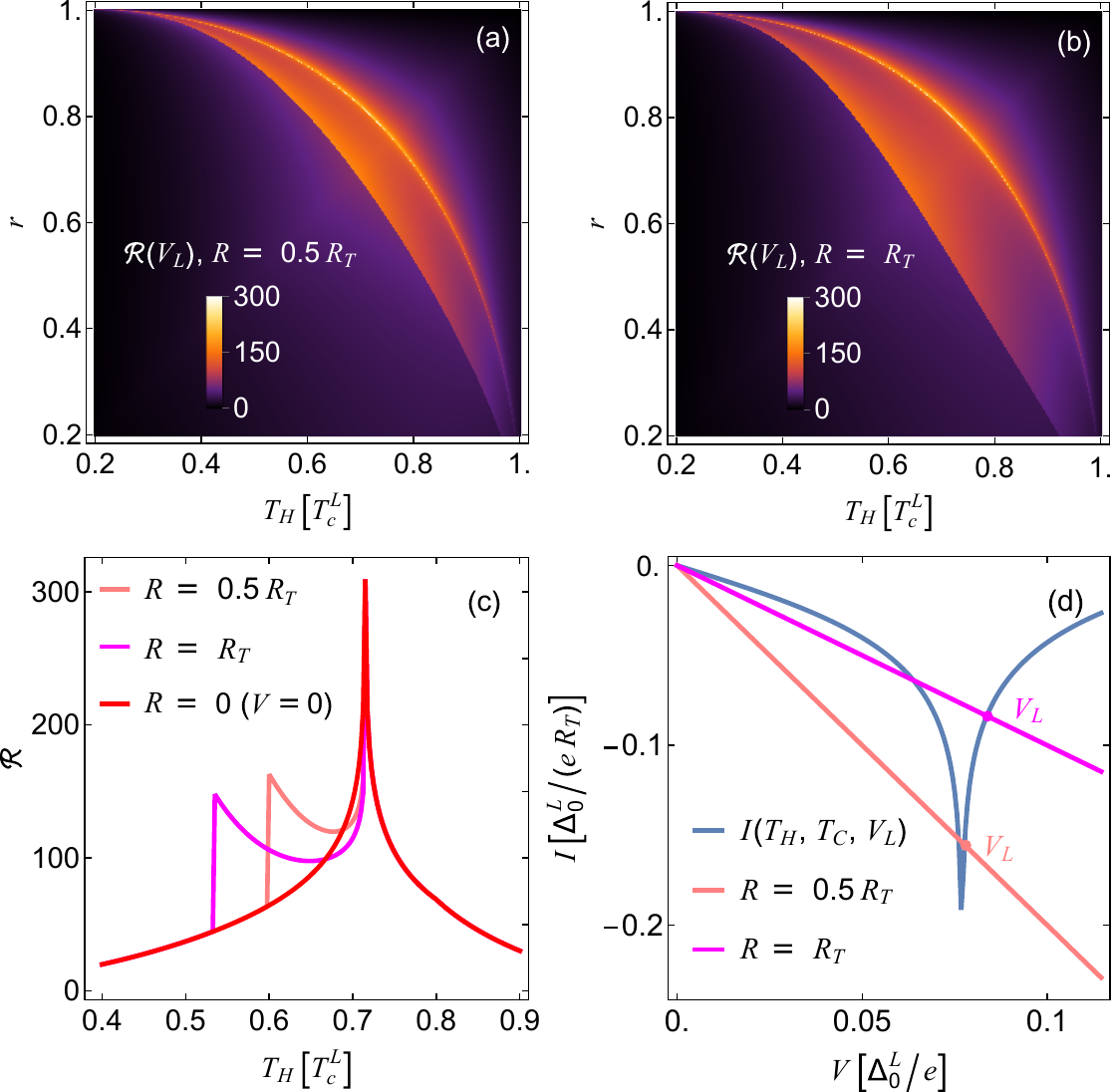}
    \caption{(a-b) Rectification efficiency $\mathcal{R}$ vs $(T_{H},r)$ at $V=V_L$ for $R=0.5R_T$ and $R=R_T$. We set $T_{C}=0.01 T_c^L$ in both cases. (c) $\mathcal{R}$ vs $T_{H}$ for $R=0.5R_T$, $R=R_T$ and $R=0$ at $r=0.8$. (d) Intersection between the load lines and $I(V,T_{H},T_{C})$ for $T_{H}=0.6T_c^L$, $T_{C}=0.01T_c^L$ and $r=0.8$. }
    \label{Figura 3}
\end{figure}
\section{Heat Pipe}
\begin{figure}[t!]
    \centering
        \includegraphics[width=\linewidth]{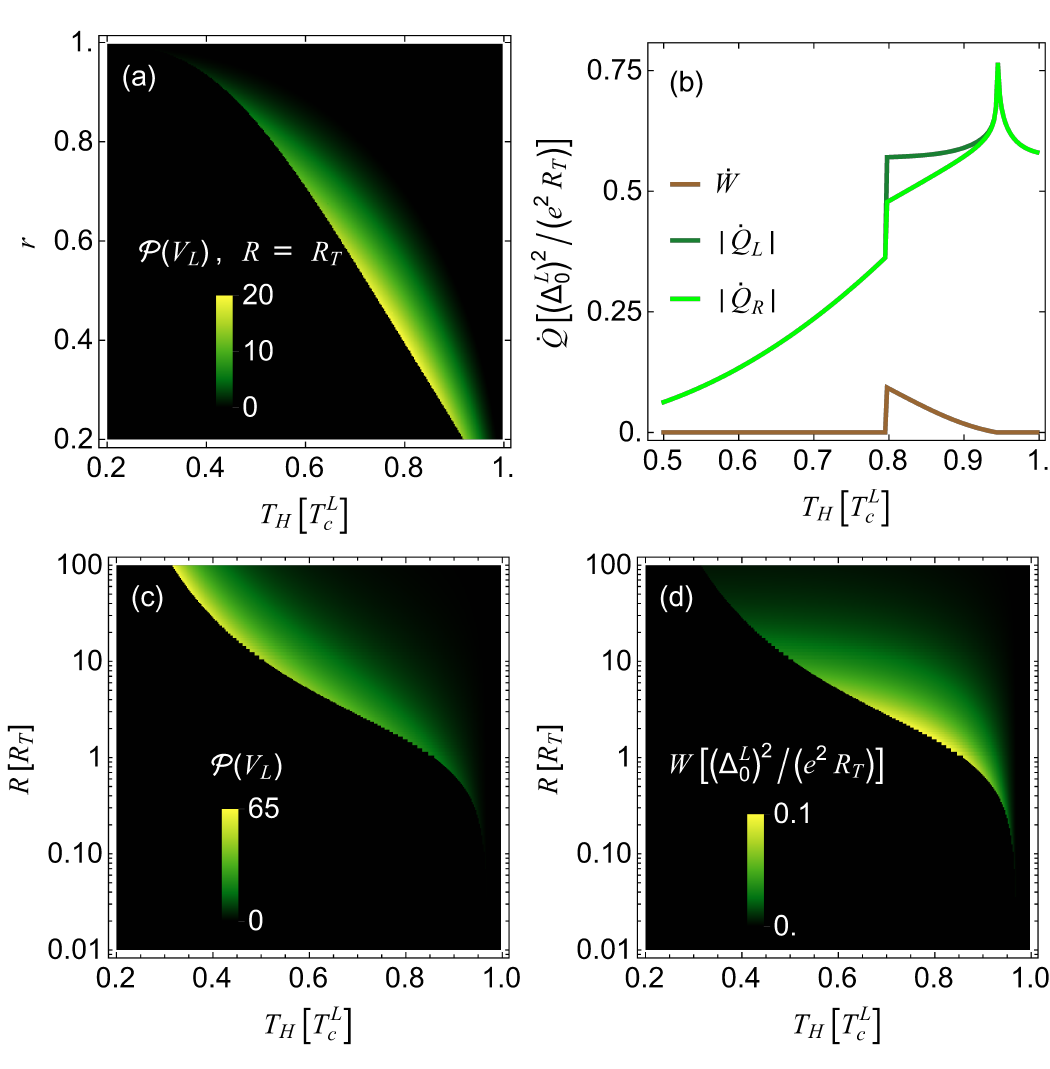}
    \caption{ (a) \(\mathcal{P}\) vs \((T_{H},r)\) for \(R=R_T\). (b) comparison between the heat current \(\dot{Q}_L\) flowing out from the L electrode, the one \(\dot{Q}_R\) coming to the right and power $\dot{W}$ dissipated on the resistor. (c) \(\mathcal{P}\) for different values of the resistor $R_L$ and $T_{H}$, calculated for $r=0.4$. (d) map of the power $\dot{W}$ dissipated on the resistor in the same conditions of (c).}
    \label{fig 4}
\end{figure}

In addition, incorporating a resistor means that a specific amount of power, \(\dot{W}=I_LV_L\), is delivered by the junction to this external component, dissipating it as Joule heating. This feature allows the potential to create a purely electronic heat pipe for cryogenic electronics. Indeed, the Joule heating is concentrated in the resistor, which can be placed at a large distance from the junction and thus, in principle, even allowing the heat to be brought outside the local cryogenic environment. This enables efficient thermal management since the heat dissipated in the cold lead is now less than the heat extracted by the hot lead, as ruled by the first law of thermodynamics.
To quantify the heat piping efficiency in this scenario, we introduce a figure of merit, \(\mathcal{P}\), defined by the equation:
\begin{equation}
    \mathcal{P} = \frac{\dot{W}}{\dot{Q}_R},
\end{equation}
Where $W$ refers to the thermoelectric power that can be transported out of the low-temperature environment via the electronic heat pipe while \(\dot{Q}_R\) is the time rate of change of the heat arriving at the cold terminal. In this way, we can reduce the necessary cooling power at the lowest temperature stage of the setup. Initially, we set the resistor to \(R = R_T\) in Fig. \ref{fig 4}(a), demonstrating that up to \(20\%\) of the thermal flux meant for the cold electrode can be redirected towards the load. Specifically, In Fig. \ref{fig 4}(b), we see how the heat fluxes and the thermoelectric power change with $T_H$ for a given value of gap ratio $r$. Intriguingly we note that thermoelectric contribution plays an important role when the temperature deviates from the optimal peak values.
In a subsequent analysis, shown in Fig. \ref{fig 4}(c), we kept the gap ratio constant while modifying resistance values. This approach revealed that increasing resistance substantially improves the system's efficiency in redirecting a part of the thermal flow to the resistor, with \(\mathcal{P}\) approaching roughly \(65\%\) in high-resistance configurations. However, this enhancement is balanced by the decrease in the amount of heat effectively redirected (thermoelectric power $\dot{W}$), greater for load in the order of the tunneling resistance, as demonstrated in Fig. \ref{fig 4}(d).
\section{Potential Experimental Implementation}
Finally, we can discuss the potential experimental implementation of this system. We aim to establish quasi-equilibrium conditions, where the quasi-particle temperature $T_{QP}$ and the phonon bath temperature $T_{Ph}$ are well-defined but can differ. In the normal state, the quasiparticle-phonon heat flux is suppressed as a power law in the sub-Kelvin regime, and it is even strongly exponentially damped in a superconductor below the critical  temperature\cite{timofeev2009recombination,giazotto2006opportunities}.  Specifically, a suitable candidate to develop these devices is Aluminum, whose excellent quality of Aluminum-oxide junctions allows the fabrication of SQUID architectures that suppress the Josephson contribution by applying an external magnetic field. To achieve a gap ratio \(r\ne1\), we can exploit the inverse proximity effect, combining Al with other superconductors or normal metals in bilayers to reduce the gap of one of the electrodes\cite{hijano2023bipolar}. For example, in the experimental evidence discussed in Ref. \cite{germanese2022bipolar}, the system studied was a Al/Cu bilayer, resulting in \(r\sim 0.35\).
\section{Conclusion}
Our research delved into the thermal transport dynamics of asymmetric \(SIS'\) Josephson junctions, emphasizing the potential role of bipolar thermoelectricity within these systems. We successfully demonstrated how this particular regime enhances both the thermal bias and the gap ratio range over which the system acts as a good thermal diode with respect to the non-thermoelectric phase of SIS'. Moreover, we revealed the potential of exploiting a load resistor to dissipate away the heat from the electrode with the larger energy gap operating as a sort of electrical heat pipe. Finally, the harvested energy can be redirected in an NIS cooling device to further increase the heat-removing capabilities of the system. This finding is especially compelling from an energy management perspective, introducing an innovative twist to conventional electrical heat piping strategies that further enhance the constrained cooling capabilities, particularly crucial at the ultra-low temperatures experienced by cold fingers. This advancement could be pivotal in improving heat control and energy harvesting within superconducting quantum technologies, marking a step forward in optimizing their operational efficiency.\\

\ack{
We acknowledge the EU’s Horizon 2020 Research and Innovation Framework Programme under Grant No. 964398 (SUPERGATE), No.
101057977 (SPECTRUM), and the PNRR MUR project
PE0000023-NQSTI for partial financial support. A.B.
acknowledges also the MUR-PRIN2022 Project NEThEQS (Grant No. 2022B9P8LN), the Royal Society through the International Exchanges between the UK and Italy (Grants No. IEC R2 192166) and CNR project QTHERMONANO. 
}

\section*{References}
\bibliographystyle{iopart-num}
\bibliography{references}
\end{document}